\documentclass[pra,twocolumn,superscriptaddress,shownopacs,amssymb]{revtex4-2}
\usepackage{graphicx,psfrag,amsmath,amssymb}

\usepackage[usenames, dvipsnames]{color}

\usepackage{hyperref} 
\hypersetup{
    colorlinks=true,
    linkcolor=Blue,
    citecolor=Blue,
    filecolor=Blue,      
    urlcolor=Blue,
    pdftitle={Multiband TDGL},
    }

\begin{document}

\title{Extracting quantum-geometric effects from Ginzburg-Landau 
theory \\ in a multiband Hubbard model}

\author{M. Iskin}
\affiliation{
Department of Physics, Ko\c{c} University, Rumelifeneri Yolu, 
34450 Sar\i yer, Istanbul, T\"urkiye
}

\date{\today}

\begin{abstract}

We first apply functional-integral approach to a multiband Hubbard model near the critical 
pairing temperature, and derive a generic effective action that is quartic in the fluctuations 
of the pairing order parameter. Then we consider time-reversal-symmetric systems with 
uniform (i.e., at both low-momentum and low-frequency) pairing fluctuations in a unit cell, 
and derive the corresponding time-dependent Ginzburg-Landau (TDGL) equation. In addition to 
the conventional intraband contribution that depends on the derivatives of the Bloch bands, 
we show that the kinetic coefficients of the TDGL equation have a geometric contribution 
that is controlled by both the quantum-metric tensor of the underlying Bloch states and 
their band-resolved quantum-metric tensors. Furthermore we show that thermodynamic 
properties such as London penetration depth, GL coherence length, GL parameter and 
upper critical magnetic field have an explicit dependence on quantum geometry. 

\end{abstract}
\maketitle

\section{Introduction}
\label{sec:intro}

In modern solid-state and condensed-matter systems, the Berry curvature tensor and 
quantum-metric tensor play vital roles in characterizing the quantum topology and 
geometry of the underlying Bloch states~\cite{provost80, berry84, resta11}. The Berry 
curvature, in particular, is crucial in calculating the Chern number, which is a 
topological invariant that quantifies the `magnetic charge' of occupied electronic 
bands in a crystal. This number determines the transport properties of some materials, 
and provides a robust way to classify and characterize the topological phases of 
matter~\cite{chiu16, bansil16}.
Another useful tool is the quantum metric, which has broad applications across various 
fields of physics, including superconductivity in multi-band 
materials~\cite{tan19, yu20, gianfrate20, tian23, yi23}. It is especially significant 
in explaining how an isolated flat band can carry finite superfluid current~\cite{torma21}.

Assuming weak interactions, time-reversal symmetry, and uniform pairing, it is possible 
to show that the quantum metric of the isolated flat band exclusively describes the 
superfluid density, allowing superconductivity to prevail~\cite{peotta15, liang17, julku16}.
Interestingly, despite the diverging band mass of its unpaired constituents and the 
smallness of onsite attractive interaction between particles, a Cooper pair can still 
acquire a finite effective mass through virtual interband transitions in the form of 
quantum metric. In addition, by solving the two-body problem in a 
multiband lattice~\cite{torma18, iskin21, iskin22}, the geometric origin of superfluid 
density can be traced all the way back to the effective mass of the lowest-lying bound 
states through an exact calculation~\cite{iskin21, iskin22}. Since the effective mass of 
Cooper pairs can be attributed to the geometric origin of superfluid 
density~\cite{iskin18a, iskin18b, huhtinen22, herzog22}, the low-energy collective modes 
(i.e., Goldstone and Leggett modes) are again influenced by the quantum metric in a similar 
way~\cite{iskin20a, iskin20b}. These peculiar findings highlight the essence of the role 
played by quantum metric in flat-band superconductivity, and they may offer a fresh 
perspective for the theory of multiband 
superconductors~\cite{torma21, rossi21, herzog22, huhtinen22}. 

In this paper, to gain further insight into the thermodynamic properties of multiband 
superconductors, we begin with a generic multiband Hubbard model and derive its effective 
quartic action in the fluctuations of the pairing order parameter near the critical 
transition temperature. Then we focus on the simpler problem, and assume that the system 
exhibits time-reversal symmetry and uniform pairing fluctuations. We demonstrate that 
the kinetic coefficients of the resulting time-dependent Ginzburg-Landau (TDGL) equation 
have a geometric interband contribution that is controlled by both the quantum-metric 
tensor of the underlying Bloch states and their band-resolved quantum-metric tensors. Finally, 
we extract the effective mass of the Cooper pairs, London penetration depth, GL coherence length, 
GL parameter, upper critical magnetic field, superfluid density and BKT transition 
temperature as examples of thermodynamic properties that depend explicitly on quantum 
geometry through a standard GL analysis.

The rest of the paper is organized as follows. In Sec.~\ref{sec:fif} we apply functional-integral
approach to the multiband Hubbard model, and derive a generic effective action that is quartic 
in the fluctuations of the pairing order parameter. In Sec.~\ref{sec:trs} we specifically consider 
time-reversal-symmetric systems with uniform pairing fluctuations, and derive and analyze the 
resultant TDGL equation. The paper ends with a brief summary of our conclusions and an outlook 
in Sec.~\ref{sec:conc}.

\section{Functional-Integral Formalism}
\label{sec:fif}

In this Section we apply imaginary-time functional path-integral approach to a generic 
multiband Hubbard model near the critical pairing temperature $T_c$, and derive an effective 
action that is quartic in the fluctuations of the pairing order 
parameter~\cite{sademelo93, stintzing97}.

\subsection{Multiband Hubbard model}
\label{sec:mHm}

Our starting Hamiltonian for a multiband Hubbard model is
$
\mathcal{H} = 
-\sum_{SS'ii' \sigma} t_{Si; S'i'}^\sigma \psi_{S i \sigma}^\dagger \psi_{S' i' \sigma}
-\sum_{Si \sigma} \mu_\sigma  \psi_{S i \sigma}^\dagger \psi_{S i \sigma}
- U \sum_{S i} 
\psi_{S i \uparrow}^\dagger \psi_{S i \downarrow}^\dagger 
\psi_{S i \downarrow} \psi_{S i \uparrow},
$
where $i$ denotes a particular unit cell in the lattice and $S$ denotes its sublattice 
sites in such a way that $t_{Si; S'i'}^\sigma$ corresponds to the hopping parameter 
from a site $S' \in i'$ to a site $S \in i$, and $\mu_\sigma$ is the chemical potential
that fixes the number of spin-$\sigma$ particles.
Thus while the kinetic energy term is quite generic, for simplicity we consider only 
attractive onsite interactions (i.e., $U \ge 0$) between $\uparrow$ and $\downarrow$ 
particles. Then we express the Hubbard Hamiltonian in reciprocal space through a 
canonical transformation 
$
\psi_{S i \sigma}^\dagger = \frac{1}{\sqrt{N_c}} \sum_\mathbf{k} 
e^{-i \mathbf{k} \cdot \mathbf{r}_{S i}} \psi_{S \mathbf{k} \sigma}^\dagger,
$
where $N_c$ is the number of unit cells in the lattice, $\mathbf{k}$ is the crystal 
momentum in the first Brillouin zone, and $\mathbf{r}_{S i}$ is the position of site 
$S \in i$. The total number of lattice sites is $N_c N_S$ where $N_S$ is the number of 
sublattice sites in a unit cell. This leads to
\begin{align}
\label{eqn:ham}
\mathcal{H} &= \sum_{S S' \mathbf{k} \sigma} \psi_{S \mathbf{k} \sigma}^\dagger 
[h_{SS'}^\sigma(\mathbf{k}) - \mu_\sigma \delta_{SS'}] \psi_{S' \mathbf{k} \sigma}
\nonumber \\
&- U \sum_{S \mathbf{k} \mathbf{k'} \mathbf{q}} 
\psi_{S \mathbf{k} \uparrow }^\dagger 
\psi_{S, -\mathbf{k}+\mathbf{q}, \downarrow}^\dagger
\psi_{S, -\mathbf{k'}+\mathbf{q}, \downarrow}
\psi_{S \mathbf{k'} \uparrow},
\end{align}
where $h_{SS'}^\sigma(\mathbf{k})$ is the matrix element of the Bloch Hamiltonian in the
sublattice basis, and $\mathbf{q}$ is the center-of-mass momentum of the incoming and 
outgoing pair of particles. Except for Sec.~\ref{sec:GLf}, we work in units of $\hbar \to 1$ 
the Planck constant.

\subsection{Effective quartic action near $T_c$}
\label{sec:eqa}

In the Grassmann functional-integral formalism~\cite{sademelo93, stintzing97}, 
the grand partition function 
$
\mathcal{Z} = \mathrm{Tr} e^{-\beta \mathcal{H}},
$ 
can be written as
$
\mathcal{Z} = \int \mathcal{D}[\psi^\dagger, \psi]e^{-\mathcal{S}},
$ 
where $\mathcal{S}$ is the associated fermionic action given by
$
\mathcal{S} = \int_0^{\beta} d\tau \big[
\sum_{S \mathbf{k} \sigma} \psi_{S \mathbf{k} \sigma}^\dagger (\tau) 
\partial_\tau \psi_{S \mathbf{k} \sigma } (\tau) + \mathcal{H}(\tau) \big].
$
Here $\tau$ is the imaginary time, and $\beta = 1/(k_B T)$ is the inverse temperature
in units of $k_B \to 1$ the Boltzmann constant.
By first introducing the usual Hubbard-Stratanovich transformation that linearizes
the interaction term at the expense of introducing a complex bosonic field 
$\Delta_S (q)$, and then integrating out the remaining fermionic degrees of freedom
that is Gaussian in the Grassmann fields, we eventually obtain
$
\mathcal{Z} = \int \mathcal{D} [\Delta^*, \Delta] e^{-\mathcal{S}_\mathrm{eff}}.
$ 
Here $\Delta_S (q)$ corresponds to the pairing order parameter, and 
\begin{align}
\mathcal{S}_\mathrm{eff} = &\beta \sum_q \frac{|\Delta_S(q)|^2}{U}  + \frac{\beta}{N_c}\sum_{S \mathbf{k}} [h_{SS}^\downarrow(\mathbf{k}) - \mu_\downarrow] 
\nonumber \\
&- \frac{1}{N_c} \mathrm{Tr} \sum_{k q} \ln [\beta \mathbf{G}^{-1}(k,q)]
\label{eqn:Seff}
\end{align}
is an effective bosonic action for the resultant pairs, where we make use of the
well-known identity
$
\ln \det \mathbf{M} = \mathrm{Tr} \ln \mathbf{M}
$
with $\mathrm{Tr}$ denoting a trace over the sublattice and particle-hole sectors,
and the $2N_S \times 2N_S$ matrix $\mathbf{G}^{-1}(k,q)$ denoting the inverse propagator.
Throughout this paper, we use the collective indices 
$
q \equiv (\mathbf{q}, \mathrm{i}\nu_\ell)
$ 
and  
$
k \equiv (\mathbf{k}, \mathrm{i}\omega_\ell),
$ 
where $\nu_\ell = 2\ell \pi/\beta$ is the bosonic Matsubara frequency and 
$\omega_\ell = (2\ell+1)\pi/\beta$ is the fermionic one. 

In order to make analytical progress with this action, we first decompose 
$
\Delta_S (q) = \Delta_S \delta_{q 0} + \Lambda_S (q)
$ 
in terms of a $q$-independent (i.e., stationary) saddle-point parameter $\Delta_S$ 
and a $q$-dependent fluctuations around it, and split the inverse propagator
\begin{align}
\mathbf{G}^{-1}(k,q) = \mathbf{G}_0^{-1}(k) + \mathbf{\Sigma}(q)
\end{align}
into two, where $\mathbf{G}_0^{-1}(k)$ depends on $\Delta_S$ and $\mathbf{\Sigma}(q)$
depends solely on $\Lambda_S (q)$. Here $\delta_{ij}$ is a Kronecker delta.
Then we reexpress the inverse propagator as
$
\mathbf{G}_0^{-1}(k)[\boldsymbol{\mathcal{I}}_{2N_S} + \mathbf{G}_0(k) \mathbf{\Sigma}(q)]
$
and expand its natural logarithm as a Taylor series in $\Lambda_S (q)$, leading to
$
\ln [\beta \mathbf{G}^{-1}(k,q)] = \ln [\beta \mathbf{G}_0^{-1}(k)] 
- \sum_{j = 1}^\infty \frac{(-1)^j}{j} [\mathbf{G}_0(k) \mathbf{\Sigma}(q)]^j.
$
In this paper we set $\Delta_S \to 0$ since we are only interested in the normal 
state properties near the critical pairing temperature when $T \to T_c$. 
This particular limit allows us to go beyond the trivial saddle-point action 
$
\mathcal{S}_0 = \frac{\beta}{N_c}\sum_{S \mathbf{k}} [h_{SS}^\downarrow(\mathbf{k}) - \mu] 
- \frac{1}{N_c} \sum_k \ln \det [\beta \mathbf{G}_0^{-1}(k)],
$
and calculate both the quadratic action $\mathcal{S}_2$ and the quartic action 
$\mathcal{S}_4$. The resultant effective action
$
\mathcal{S}_\mathrm{eff} \approx \mathcal{S}_0 + \mathcal{S}_2 + \mathcal{S}_4
$
is expected to be qualitatively accurate in describing the low-energy physics at all 
coupling strengths.

When the saddle-point order parameter $\Delta_S \to 0$ is trivial, the saddle-point 
component of the inverse propagator
$
\mathbf{G}_0^{-1}(k) = \left[
\begin{array}{cc}
 (\mathbf{G}_0^{11})^{-1}(k) & 0 \\
  0 & (\mathbf{G}_0^{22})^{-1}(k)
\end{array}
\right]
$
becomes diagonal in the particle-hole sector, where its matrix element
$
(G_0^{11})^{-1}_{SS'}(k) = (\mathrm{i}\omega_\ell + \mu_\uparrow) \delta_{SS'} 
- h_{SS'}^\uparrow(\mathbf{k})
$
describes a spin-$\uparrow$ particle in the sublattice sector. Similarly
$
(G_0^{22})^{-1}_{SS'}(k)  = (\mathrm{i}\omega_\ell - \mu_\downarrow) \delta_{SS'} 
+ h_{SS'}^{\downarrow *}(-\mathbf{k})
$
describes a spin-$\downarrow$ hole. On the other hand the fluctuation component
$
\boldsymbol{\Sigma}(q) = 
\left[
\begin{array}{cc}
 0 & \boldsymbol{\Sigma}^{12}(q) \\
  \boldsymbol{\Sigma}^{21}(q)  & 0
\end{array}
\right]
$
is off-diagonal in the particle-hole sector by definition, where
$
\Sigma_{SS'}^{12}(q) = \Lambda_S(q) \delta_{SS'}
$
and
$
\Sigma_{SS'}^{21}(q) = \Lambda_S^*(-q) \delta_{SS'}
$
are diagonal in the sublattice sector due to the onsite interactions considered
in this paper. For convenience we represent the matrix elements of the 
non-interacting propagators $\mathbf{G}_0^{11}(k)$ and $\mathbf{G}_0^{22}(k)$ as
\begin{align}
\label{eqn:G11}
G^{11}_{SS'}(k) &= \sum_{n} \frac{n_{S \mathbf{k} \uparrow} n_{S' \mathbf{k} \uparrow}^*}
{\mathrm{i}\omega_\ell - \xi_{n \mathbf{k} \uparrow }},
\\
\label{eqn:G22}
G^{22}_{SS'}(k) &= \sum_{n} \frac{n_{S,-\mathbf{k}, \downarrow}^* n_{S',-\mathbf{k}, \downarrow}}
{\mathrm{i}\omega_\ell + \xi_{n, -\mathbf{k}, \downarrow}},
\end{align}
where $n_{S \mathbf{k} \sigma} = \langle S | n \mathbf{k} \sigma \rangle$ is the 
projection of the periodic part of the Bloch state onto sublattice $S$, and
$
\xi_{n \mathbf{k} \sigma} = \varepsilon_{n \mathbf{k} \sigma} - \mu_\sigma
$
is associated with the corresponding Bloch band $\varepsilon_{n \mathbf{k} \sigma}$.
They are in such a way that
$
\sum_{S'} h_{SS'}^\sigma(\mathbf{k}) n_{S' \mathbf{k} \sigma} 
= \varepsilon_{n \mathbf{k} \sigma} n_{S \mathbf{k} \sigma},
$
and the propagators satisfy 
$
\mathbf{G}_0^{11}(k) (\mathbf{G}_0^{11})^{-1}(k) = \boldsymbol{\mathcal{I}}_{N_S}
$
and 
$
\mathbf{G}_0^{22}(k) (\mathbf{G}_0^{22})^{-1}(k) = \boldsymbol{\mathcal{I}}_{N_S},
$ 
where $\boldsymbol{\mathcal{I}}_{N_S}$ is an $N_S \times N_S$ identity matrix.

\begin{figure}[!htb]
    \centering
    \vskip 5mm
    \includegraphics[width=0.5\columnwidth]{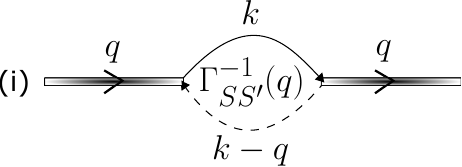}
    \vskip 5mm
    \includegraphics[width=0.6\columnwidth]{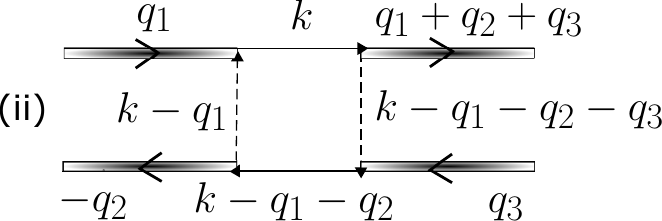}
    \caption{
    Diagrammatic representation of the fluctuation contributions that give rise to 
    (i) quadratic action and (ii) quartic action. 
    While the pair propagators are shown as colored bars, the particle and hole propagators 
    are shown as solid and dashed lines, respectively.
    }
    \label{fig:Fdiag} 
\end{figure}

Using Eqs.~(\ref{eqn:G11}) and (\ref{eqn:G22}) in the quadratic action 
$
\mathcal{S}_2 = \frac{\beta}{U} \sum_{S q} |\Lambda_S (q)|^2
+ \frac{1}{2 N_c} \mathrm{Tr} \sum_{kq} \mathbf{G}_0(k) \mathbf{\Sigma}(q) 
\mathbf{G_0}(k-q) \mathbf{\Sigma}(-q),
$
and after evaluating the trace, we obtain
\begin{align}
\label{eqn:S2}
\mathcal{S}_2 = \beta \sum_{S S' q} \Lambda_S^*(q) \Gamma^{-1}_{SS'}(q) \Lambda_{S'} (q),
\end{align}
where 
$
\Gamma^{-1}_{SS'}(q) = \delta_{SS'}/U 
+ \frac{1}{\beta N_c} \sum_k G^{11}_{SS'}(k) G^{22}_{S'S}(k-q)  
$
is the matrix element of the inverse pair fluctuation propagator $\boldsymbol{\Gamma}^{-1}(q)$.
The corresponding Feynman diagram is sketched in Fig.~\ref{fig:Fdiag}(i). 
By evaluating the Matsubara summation over $\omega_\ell$, we find
\begin{align}
\Gamma_{SS'}^{-1}(q) = \frac{\delta_{SS'}}{U} &+ \frac{1}{2 N_c} \sum_{nm\mathbf{k}} 
\frac{\mathcal{X}_{n \mathbf{k} \uparrow} + \mathcal{X}_{m, -\mathbf{k}+\mathbf{q}, \downarrow}}
{\mathrm{i}\nu_\ell - \xi_{n \mathbf{k} \uparrow} -  \xi_{m,-\mathbf{k}+\mathbf{q}, \downarrow}}
\nonumber \\ 
& \times n_{S \mathbf{k} \uparrow} n_{S' \mathbf{k} \uparrow}^* 
m_{S', -\mathbf{k}+\mathbf{q}, \downarrow}^* m_{S, -\mathbf{k}+\mathbf{q}, \downarrow},
\label{eqn:GammaSS}
\end{align}
where
$
\mathcal{X}_{n\mathbf{k}\sigma} = \tanh(\beta \xi_{n\mathbf{k}\sigma}/2) 
= 1 - 2f(\xi_{n\mathbf{k}\sigma})
$
is the usual thermal factor with $f(\varepsilon) = 1/(e^{\beta \varepsilon}+1)$ the
Fermi-Dirac distribution. 

There are two important remarks about Eq.~(\ref{eqn:GammaSS}).
First we note that the generalized Thouless condition 
$
\det \boldsymbol{\Gamma}^{-1}(\mathbf{0}, 0) = 0
$
determines $T_c$ in a multiband Hubbard model~\cite{umucalilar16}. This is in such a 
way that 
$
T_c = \max\{ T_{c_1}, T_{c_2}, \cdots, T_{c_{N_S}}\},
$
where $T_{c_j}$ is determined by setting the $j$th eigenvalue of 
$\boldsymbol{\Gamma}^{-1}(\mathbf{0}, 0)$ to $0$. In calculating $T_c$ one needs to 
determine $\mu_\sigma$ self-consistently through the number equation 
$
\mathcal{N}_\sigma = - \partial \Omega / \partial \mu_\sigma
\approx \mathcal{N}^\sigma_0 + \mathcal{N}^\sigma_2,
$
where $\Omega = - \ln \mathcal{Z}/\beta$ is the grand potential. This leads to a
saddle-point contribution 
$
\mathcal{N}^\sigma_0 = \sum_{n \mathbf{k}} (1 - \mathcal{X}_{n\mathbf{k}\sigma})/2
$
that originates from $\partial (\mathcal{S}_0/\beta) / \partial \mu_\sigma$ and a
fluctuation contribution
$
\mathcal{N}^\sigma_2 = - \frac{1}{\beta} \sum_{q} \partial\{
\ln\det [\boldsymbol{\Gamma}^{-1}(q)/\beta]\}/\partial \mu_\sigma
$
that is due to $\mathcal{S}_2$ after performing the Gaussian integration over the
bosonic fluctuation fields. The saddle-point contribution by itself gradually fails to 
describe the low-energy physics with increasing coupling strengths, i.e., even at a 
qualitative level, and taking proper account of the fluctuation contribution is known 
to be a nontrivial yet crucial task in describing the strong-coupling 
limit~\cite{nsr85, sademelo93}.
Second we are pleased to confirm that Eq.~(\ref{eqn:GammaSS}) and the associated Thouless 
condition are in perfect agreement with the self-consistency relation that determines 
the exact energy $E_{2b}(\mathbf{q})$ of the two-body bound states~\cite{iskin21, iskin22},
after setting the thermal factors $\mathcal{X}_{n \mathbf{k} \sigma} \to 1$ in 
$
\boldsymbol{\Gamma}^{-1}(\mathbf{q}, \mathrm{i}\nu_\ell \to \omega + \mathrm{i}0^+)
$ 
together with the substitution $\omega + 2\mu \to E_{2b}(\mathbf{q})$. 

Similarly, using Eqs.~(\ref{eqn:G11}) and (\ref{eqn:G22}) in the quartic action 
$
\mathcal{S}_4 = \frac{1}{4 N_c} \mathrm{Tr} \sum_{kq_1 q_2 q_3 q_4} 
\mathbf{G}_0(k) \mathbf{\Sigma}(q_1) 
\mathbf{G_0}(k-q_1) \mathbf{\Sigma}(q_2)
\mathbf{G_0}(k-q_1-q_2) \mathbf{\Sigma}(q_3)
\mathbf{G_0}(k-q_1-q_2-q_3) \mathbf{\Sigma}(-q_1-q_2-q_3),
$
and after evaluating the trace, we obtain
\begin{align}
\mathcal{S}_4 = \frac{\beta}{2} \sum_{\substack{S S' S'' S''' \\ q_1 q_2 q_3}} 
& b_{SS'S''S'''}(q_1,q_2,q_3) \Lambda_S^*(q_1+q_2+q_3)
\nonumber \\ 
& \times \Lambda_{S'}(q_1) \Lambda_{S''}^*(-q_2) \Lambda_{S'''}(q_3).
\label{eqn:S4}
\end{align}
The corresponding Feynman diagram is sketched in Fig.~\ref{fig:Fdiag}(ii). 
While the $q$-dependence of the prefactor
$
b_{SS'S''S'''}(q_1,q_2,q_3) = \frac{1}{\beta N_c} \sum_k G^{11}_{SS'}(k) G^{22}_{S'S''}(k-q_1)
G^{11}_{S''S'''}(k-q_1-q_2) G^{22}_{S'''S}(k-q_1-q_2-q_3)
$
is quite complicated in general, we approximate it with its stationary value at 
zero momentum and zero frequency (i.e., at $q_1 = q_2 = q_3 = 0$), which is a 
standard practice in basic BCS-BEC crossover theories~\cite{sademelo93, stintzing97}.
By evaluating the Matsubara summation over $\omega_\ell$, we find
\begin{widetext}
\begin{align}
b_{SS'S''S'''}(0,0,0) = \frac{1}{2N_c} & \sum_{n_1 n_2 n_3 n_4 \mathbf{k}}
\frac{{n_1}_{S \mathbf{k} \uparrow} {n_1}_{S' \mathbf{k} \uparrow}^*
{n_2}_{S', -\mathbf{k}, \downarrow}^* {n_2}_{S'', -\mathbf{k}, \downarrow}
{n_3}_{S'' \mathbf{k} \uparrow} {n_3}_{S''' \mathbf{k} \uparrow}^* 
{n_4}_{S''', -\mathbf{k}, \downarrow}^* {n_4}_{S, -\mathbf{k}, \downarrow}}
{(\xi_{n_1 \mathbf{k} \uparrow } + \xi_{n_2, -\mathbf{k}, \downarrow})
(\xi_{n_3 \mathbf{k} \uparrow } + \xi_{n_4, -\mathbf{k}, \downarrow})}
\nonumber \\
&\times \left[
\frac{\mathcal{X}_{n_1 \mathbf{k} \uparrow} + \mathcal{X}_{n_4, -\mathbf{k}, \downarrow}}
{\xi_{n_1 \mathbf{k} \uparrow} + \xi_{n_4, -\mathbf{k}, \downarrow}}
+ \frac{\mathcal{X}_{n_3 \mathbf{k} \uparrow} + \mathcal{X}_{n_2, -\mathbf{k}, \downarrow}}
{\xi_{n_3 \mathbf{k} \uparrow} + \xi_{n_2, -\mathbf{k}, \downarrow}}
- \frac{\mathcal{X}_{n_1 \mathbf{k} \uparrow} - \mathcal{X}_{n_3 \mathbf{k} \uparrow}}
{\xi_{n_1 \mathbf{k} \uparrow} - \xi_{n_3 \mathbf{k} \uparrow}}
- \frac{\mathcal{X}_{n_2, -\mathbf{k}, \downarrow} - \mathcal{X}_{n_4, -\mathbf{k}, \downarrow}}
{\xi_{n_2, -\mathbf{k}, \downarrow} - \xi_{n_4, -\mathbf{k}, \downarrow}}
\right].
\label{eqn:bS0}
\end{align}
\end{widetext}
As discussed in Sec.~\ref{sec:GLf}, this parameter characterizes the interactions between Cooper
pairs. Note that a derivative is implied by the last two terms when the summation indices $n_1$ 
and $n_3$ or similarly $n_2$ and $n_4$ refer to the same Bloch band,  i.e.,
$
d[\tanh(\beta \varepsilon/2)]/d\varepsilon 
= \frac{\beta}{2} \mathrm{sech}^2(\beta \varepsilon/2).
$
Thus Eqs.~(\ref{eqn:S2}), (\ref{eqn:GammaSS}), (\ref{eqn:S4}) and (\ref{eqn:bS0}) all together 
constitute our aforementioned action 
$
\mathcal{S}_\mathrm{eff} \approx \mathcal{S}_0 + \mathcal{S}_2 + \mathcal{S}_4
$ 
near $T_c$ for a generic multiband Hubbard model.

\section{Time-reversal-symmetric systems with uniform pairing fluctuations}
\label{sec:trs}

In order to make further analytical progress, here we consider Hubbard models that 
manifest time-reversal symmetry, and set
$
n_{S, -\mathbf{k}, \downarrow}^* = n_{S \mathbf{k} \uparrow} \equiv n_{S \mathbf{k}} 
$
and 
$
\xi_{n, -\mathbf{k}, \downarrow} = \xi_{n \mathbf{k} \uparrow} \equiv \xi_{n \mathbf{k}}.
$
Furthermore we limit our discussion to those systems where the low-$q$ pairing 
fluctuations (i.e., at both low momentum and low frequency) are uniform in a unit cell,
and set $\Lambda_S(q) = \Lambda_0(q)$ for all sublattices. 
For instance Mielke checkerboard and Kagome lattices may satisfy this condition 
because of their inversion symmetry as they satisfy the analogous condition in the 
two-body problem~\cite{iskin21, iskin22}. The low-energy physics turns out 
to be very transparent in this particular limit as discussed next.

\subsection{Effective quartic action}
\label{sec:ea}

When both time-reversal-symmetry and uniform-pairing-fluctuation conditions 
met simultaneously, Eq.~(\ref{eqn:S2}) can be written as
$
\mathcal{S}_2 = \beta \sum_{q} \Gamma^{-1}_0(q) |\Lambda_0 (q)|^2
$
and Eq.~(\ref{eqn:S4}) can be written as
$
\mathcal{S}_4 = \frac{\beta b_0}{2}
\sum_{q_1 q_2 q_3}
\Lambda_0^*(q_1+q_2+q_3)\Lambda_0(q_1) \Lambda_0^*(-q_2) \Lambda_0(q_3).
$
Using the completeness relation
$
\sum_S |S \mathbf{k} \rangle \langle S \mathbf{k} | = \boldsymbol{\boldsymbol{\mathcal{I}}}_{N_S}
$
for the sublattice basis, Eqs.~(\ref{eqn:GammaSS}) and (\ref{eqn:bS0}) reduce to
\begin{align}
\Gamma^{-1}_0(q) &= \frac{N_S}{U} + \frac{1}{2 N_c} \sum_{nm\mathbf{k}}
\frac{
\mathcal{X}_{n \mathbf{k}} + \mathcal{X}_{m, \mathbf{k-q}} }
{\mathrm{i}\nu_\ell - \xi_{n\mathbf{k}} - \xi_{m, \mathbf{k-q}}}
|\langle n_\mathbf{k} | m_\mathbf{k-q} \rangle|^2,
\label{eqn:Gamma0}
\\
b_0 &= \frac{1}{N_c} \sum_{n\mathbf{k}} \left(
\frac{\mathcal{X}_{n \mathbf{k}}}{4\xi_{n\mathbf{k}}^3}
- \frac{\beta \mathcal{Y}_{n\mathbf{k}}}{8\xi_{n\mathbf{k}}^2}
\right),
\label{eqn:b0}
\end{align}
where 
$
\mathcal{Y}_{n\mathbf{k}} = \mathrm{sech^2}(\beta \xi_{n\mathbf{k}}/2)
$
is another thermal factor. In the presence of a single band, Eqs.~(\ref{eqn:Gamma0}) 
and (\ref{eqn:b0}) recover the usual results~\cite{sademelo93, stintzing97} after 
setting $N_S \to 1$ and 
$
\langle n_\mathbf{k} | m_\mathbf{k-q} \rangle \to \delta_{nm}
$
since $n_{S \mathbf{k}} = 1$ is trivial for every $\mathbf{k}$ state. Furthermore, in the 
presence of two bands, all of our results here and below are in full agreement with the 
recent literature on spin-orbit-coupled Fermi gases. There the helicity bands of a 
continuum model play exactly the same role as the Bloch bands of the lattice model as 
they also manifest time-reversal symmetry and naturally exhibit uniform pairing 
fluctuations in the spin sector~\cite{iskin18a, iskin18b}. Next we show that the Bloch factor
$
|\langle n_\mathbf{k} | m_\mathbf{k-q} \rangle|^2
$
appearing in Eq.~(\ref{eqn:Gamma0}) is responsible for the appearance of quantum-geometric 
terms in the presence of two or more Bloch bands.

\subsection{Low-momentum and low-frequency expansion}
\label{sec:lee}

In the low-momentum regime, we expand the static part of the inverse pair propagator 
$
\Gamma^{-1}_0(\mathbf{q}, 0) = a(T) + \frac{1}{2} \sum_{ij} c_{ij} q_i q_j + \cdots
$
as a Taylor series in $\mathbf{q}$~\cite{sademelo93, stintzing97}, and split the 
kinetic coefficient
$
c_{ij} = c_{ij}^\mathrm{intra} + c_{ij}^\mathrm{inter}
$
into two contributions depending on whether the intraband or interband transition processes 
are involved. Here $q_i$ refers to components of $\mathbf{q} = (q_x, q_y, q_z)$.
A compact way to express the resultant expansion coefficients is
\begin{align}
\label{eqn:aT}
a(T) &= \frac{N_S}{U} - \frac{1}{N_c} \sum_{n \mathbf{k}} 
\frac{\mathcal{X}_{n\mathbf{k}}}{2\xi_{n\mathbf{k}}},
\\
c_{ij}^\mathrm{intra} &= \frac{1}{N_c} \sum_{n \mathbf{k}} \bigg[\left( 
\frac{\mathcal{X}_{n\mathbf{k}}}{8\xi_{n\mathbf{k}}^2} 
- \frac{\beta \mathcal{Y}_{n\mathbf{k}}}{16\xi_{n\mathbf{k}}} 
\right) \ddot{\xi}_{n\mathbf{k}}^{ij} 
\nonumber \\
&\;\;\;\;\;\;\;\;\;\;\;\;\;\;\;\;
+\beta^2 \frac{\mathcal{X}_{n\mathbf{k}}\mathcal{Y}_{n\mathbf{k}}}{16 \xi_{n\mathbf{k}}}
\dot{\xi}_{n\mathbf{k}}^i \dot{\xi}_{n\mathbf{k}}^j
\bigg],
\label{eqn:cintra}
\\
c_{ij}^\mathrm{inter} &= \frac{1}{N_c} \sum_{n \mathbf{k}}
\frac{\mathcal{X}_{n\mathbf{k}}}{2\xi_{n\mathbf{k}}} g^{n\mathbf{k}}_{ij}
-  \frac{1}{2N_c} \sum_{n \ne m, \mathbf{k}} \frac{\mathcal{X}_{n\mathbf{k}} + \mathcal{X}_{m\mathbf{k}}}
{\xi_{n\mathbf{k}}+\xi_{m\mathbf{k}}} g^{nm\mathbf{k}}_{ij},
\label{eqn:cinter}
\end{align}
where the intraband contribution depends on the derivatives
$
\dot{\xi}_{n\mathbf{k}}^i \equiv \partial \xi_{n\mathbf{k}} / \partial k_i
$
and
$
\ddot{\xi}_{n\mathbf{k}}^{ij} \equiv \partial^2 \xi_{n\mathbf{k}} / (\partial k_i \partial k_j)
$
of the Bloch bands, and the interband contribution depends on the derivatives of 
the Bloch states through $g^{n\mathbf{k}}_{ij}$ and $g^{nm\mathbf{k}}_{ij}$ (defined 
further below). We note that the Thouless condition takes the usual form $a(T_c) = 0$, 
and determines $T_c$. For this reason we expand the zeroth-order coefficient as
$
a(T) = - a_0 \epsilon (T),
$ 
where 
$
a_0 = T_c \left[ \partial a(T)/\partial T \right]_{T_c}
$
and 
$
\epsilon (T) = (1 - T/T_c).
$
This leads to
\begin{align}
\label{eqn:a0}
a_0 = \frac{1}{2N_c} \sum_{n \mathbf{k}} \bigg[
\frac{\mathcal{Y}_{n \mathbf{k}}}{2T_c} 
+ \frac{\partial \mu}{\partial T} \bigg( \frac{\mathcal{Y}_{n \mathbf{k}}}{2\xi_{n \mathbf{k}}} 
- T_c \frac{\mathcal{X}_{n \mathbf{k}}}{\xi_{n \mathbf{k}}^2}  \bigg)
\bigg],
\end{align}
where the thermal factors are evaluated at $T_c$, and $\mu$ needs to be determined 
self-consistently via the number equation
$
\mathcal{N} = \sum_\sigma \mathcal{N}^\sigma \approx \mathcal{N}_0 + \mathcal{N}_2.
$
Here 
$
\mathcal{N}_0 = \sum_{n \mathbf{k}} (1 - \mathcal{X}_{n\mathbf{k}})
$
is the saddle-point contribution and 
$
\mathcal{N}_2 = \frac{1}{\beta} \sum_{q} \partial\{\ln[\beta \Gamma_0(q)]\}/\partial \mu
$
comes from the fluctuations~\cite{sademelo93, nsr85}.

The kinetic coefficient Eq.~(\ref{eqn:cintra}) is nothing but a summation over the usual 
single-band coefficient in disguise~\cite{iskin18b, sademelo93}. This can be revealed by 
performing an integration by parts
$
\sum_\mathbf{k} \mathcal{X}_{n\mathbf{k}} \dot{\xi}_{n\mathbf{k}}^i \dot{\xi}_{n\mathbf{k}}^j 
/ \xi_{n\mathbf{k}}^3
= \beta \sum_\mathbf{k} \mathcal{Y}_{n\mathbf{k}} \dot{\xi}_{n\mathbf{k}}^i \dot{\xi}_{n\mathbf{k}}^j
/ (4\xi_{n\mathbf{k}}^2)
+ \sum_\mathbf{k} \mathcal{X}_{n\mathbf{k}} \ddot{\xi}_{n\mathbf{k}}^{ij} / (2\xi_{n\mathbf{k}}^2)
$
that is followed by yet another integration by parts
$\sum_\mathbf{k} \mathcal{Y}_{n\mathbf{k}} \dot{\xi}_{n\mathbf{k}}^i \dot{\xi}_{n\mathbf{k}}^j
/ \xi_{n\mathbf{k}}^2
= -\beta \sum_\mathbf{k} \mathcal{X}_{n\mathbf{k}} \mathcal{Y}_{n\mathbf{k}} 
\dot{\xi}_{n\mathbf{k}}^i \dot{\xi}_{n\mathbf{k}}^j / \xi_{n\mathbf{k}}
+ \sum_\mathbf{k} \mathcal{Y}_{n\mathbf{k}} \ddot{\xi}_{n\mathbf{k}}^{ij} / \xi_{n\mathbf{k}}^2
$
~\cite{iskin18c}. Thus $c_{ij}^\mathrm{intra}$ is precisely the so-called conventional 
contribution arising from the intraband processes.
On the other hand the kinetic coefficient Eq.~(\ref{eqn:cinter}) depends on both the 
quantum-metric tensor $g_{ij}^{n\mathbf{k}}$ of the $n$th Bloch band and its band-resolved 
quantum-metric tensor $g_{ij}^{n m\mathbf{k}}$ defined as
\begin{align}
\label{eqn:gijn}
g_{ij}^{n\mathbf{k}} &= 2\mathrm{Re} \left[
\langle \dot{n}_\mathbf{k}^i |
\big( \boldsymbol{\mathcal{I}}_{N_S}- | n_\mathbf{k} \rangle \langle n_\mathbf{k}| \big) 
| \dot{n}_\mathbf{k}^j \rangle 
\right],
\\
\label{eqn:gijnm}
g_{ij}^{n m\mathbf{k}} &= 2\mathrm{Re} \left[ 
\langle \dot{n}_\mathbf{k}^i | m_\mathbf{k} \rangle
\langle m_\mathbf{k} | \dot{n}_\mathbf{k}^j \rangle
\right],
\end{align}
where $\mathrm{Re}$ denotes the real part,
$
g_{ij}^{n\mathbf{k}}  = \sum_{n\ne m} g_{ij}^{n m\mathbf{k}},
$
and 
$
| \dot{n}_\mathbf{k}^i \rangle \equiv \partial (|n_{\mathbf{k}} \rangle)/\partial k_i.
$
For this reason $c_{ij}^\mathrm{inter}$ is the so-called geometric contribution arising
from the virtual interband processes~\footnote{
According to Ref.~\cite{huhtinen22}, the quantum metric is guaranteed to the 
so-called minimal quantum metric, when the orbitals of the lattice model 
are fixed by symmetries at high-symmetry points. This symmetry requirement must 
be equivalent to our condition on uniform pairing fluctuations. For instance 
inversion symmetry guarantees this in the case of Mielke checkerboard and Kagome 
lattices~\cite{iskin21, iskin22}.
}. 
The geometric terms appear only in a multiband Hubbard 
model, and their physical origin can be traced back to the Bloch factor appearing in 
Eq.~(\ref{eqn:Gamma0}). For instance, by expanding the Bloch state
$
| m_\mathbf{k-q} \rangle = |m_\mathbf{k} - \sum_i \dot{m}_\mathbf{k}^i q_i + 
\frac{1}{2} \sum_{ij} \ddot{m}_\mathbf{k}^{ij} q_i q_j + \cdots \rangle
$
as a Taylor series in $\mathbf{q}$, one can easily show that
$
|\langle n_\mathbf{k} | m_\mathbf{k-q} \rangle|^2 
= \delta_{nm} - \frac{1}{2} \sum_{ij} [g_{ij}^{n\mathbf{k}} \delta_{nm} 
+ g_{ij}^{nm\mathbf{k}}(\delta_{nm} - 1)] q_i q_j + \cdots
$
up to second-order in $\mathbf{q}$. Here the first derivative of the ortho-normalization 
condition
$
\langle n_\mathbf{k} | m_\mathbf{k} \rangle = \delta_{nm}
$ 
shows that the first-order coefficients vanish, where we also use its second derivative 
to obtain the final form. 

In order to extract the low-frequency dependence of the inverse pair 
propagator, we expand 
$
Q(\mathrm{i} \nu_\ell) = \Gamma^{-1}_0(\mathbf{0}, \mathrm{i} \nu_\ell) 
- \Gamma^{-1}_0(\mathbf{0},0)
$
in powers of $\omega$ after the analytic continuation 
$
\mathrm{i} \nu_\ell \to \omega + \mathrm{i}0^+
$
~\cite{sademelo93}. Using the Cauchy principal value
$
1 / (x \pm \mathrm{i} 0^+) = \mathcal{P}(1/x) \mp \mathrm{i} \pi \delta(x),
$
we expand $Q(\omega) = -d \omega + \cdots$, and obtain
$
d = - \partial Q (\omega)/\partial \omega |_{\omega = 0}
$
as
\begin{align}
d &=  \frac{1}{N_c} \sum_{n \mathbf{k}} \frac{\mathcal{X}_{n\mathbf{k}}}{4\xi_{n\mathbf{k}}^2}
+ \frac{\mathrm{i} \pi \beta}{8 N_c} \sum_n D_n(\mu) \theta_\mu.
\end{align}
Here $\delta(x)$ is the Dirac-delta distribution,
$
D_n(\varepsilon) = \sum_\mathbf{k} \delta(\varepsilon - \varepsilon_{n\mathbf{k}})
$
is the density of states for the $n$th Bloch band, and 
$
\theta_\varepsilon = \theta(\varepsilon - \min\{\varepsilon_{n\mathbf{k}}\})
\theta(\max\{\varepsilon_{n\mathbf{k}}\} - \varepsilon)
$
with $\theta(x)$ the Heaviside-step function. Thus the coefficient $d$ has a positive imaginary 
part when $\mu$ lies within any one of the Bloch bands. 
Having derived the low-momentum and low-frequency expansion coefficients $a(T)$, $b_0$, 
$c_{ij}$ and $d$, we are ready to discuss the underlying GL theory
and extract some quantum-geometric effects from it.

\subsection{Ginzburg-Landau (GL) functional}
\label{sec:GLf}

For slowly-varying order parameter $\Lambda_0(x)$ in space and time near $T_c$, 
the TDGL equation 
\begin{align}
\bigg[
a(T) + b_0 |\Lambda_0(x)|^2 - \sum_{ij} \frac{c_{ij}}{2} \nabla_i \nabla_j 
\bigg] \Lambda_0(x) 
= \mathrm{i} \hbar d \frac{\partial \Lambda_0(x)}{\partial t}
\end{align}
is obtained by Fourier transforming the minimum action condition 
$
\partial \mathcal{S}_{\mathrm{eff}}/\partial \Lambda_0^*(x) = 0
$ 
to real-space representation $x \equiv (\mathbf{x}, t)$, where $\nabla_i$ refers to 
the components of the gradient operator~\cite{sademelo93, stintzing97}.
Here $b_0$, $c_{ij}$ and $d$ are evaluated at $T = T_c$.
When the imaginary part of $d$ is nonzero, e.g., in the weak coupling limit, this equation 
suggests that the dynamics of $\Lambda_0(x)$ is overdamped as a reflection of the continuum 
of fermionic excitations into which a Cooper pair can decay. On the other hand $\Lambda_0(x)$ is 
propagating when $d$ becomes purely real, e.g., in the strong-coupling limit away from
half-filling. In the latter case one can scale the order parameter as 
$\Lambda_B(x) = \sqrt{d} \Lambda_0(x)$, and conclude that 
\begin{align}
\big(M_B^{-1}\big)_{ij} = \frac{c_{ij}}{\hbar^2 d}
\end{align}
corresponds to the matrix elements of the inverse effective-mass tensor $\mathbf{M}_B^{-1}$ 
of the Cooper pairs. Thus the quantum-metric tensor directly controls the inverse 
effective-mass tensor of the Cooper pairs~\cite{iskin18b}. In addition one also identifies
$\mu_B(T) = -a(T)/d$ as the effective pair chemical potential and $U_{BB} = b_0/d^2$ as the 
effective pair-pair repulsion. We are pleased to confirm that the former conclusion coincides 
precisely with the inverse effective-mass tensor of the lowest-lying two-body bound 
states~\cite{iskin21, iskin22}, after setting the thermal factors 
$
\mathcal{X}_{n \mathbf{k}} \to 1
$ 
and 
$
\mathcal{Y}_{n \mathbf{k}} \to 0
$ 
together with the substitution $2\mu \to E_{2b}(\mathbf{0})$ in the remaining terms. 
Here $E_{2b}(\mathbf{0})$ is the energy of the lowest-lying two-body bound state at 
$\mathbf{q} = \mathbf{0}$, and its self-consistency relation coincides with the Thouless 
condition $a(T_c) = 0$ under the same settings~\cite{iskin21, iskin22}.

To study thermodynamic properties next we take a time-independent order parameter 
$
\Lambda_0(x) \equiv \Lambda_0(\mathbf{x}),
$
and consider an isotropic kinetic coefficient $c_{ij} = c \delta_{ij}$ for its simplicity. 
Then we scale the order parameter as 
$
\Psi (\mathbf{x}) = \sqrt{m_0 c/\hbar^2} \Lambda (\mathbf{x}),
$ 
and substitute $\nabla_i \to \nabla_i - 2\mathrm{i}q_0 A_i/(\hbar c_0)$ to introduce an 
external magnetic field 
$
\mathbf{H} = \boldsymbol{\nabla} \times \mathbf{A},
$ 
where $q_0$ is the charge of the particles, $c_0$ is the speed of light, $m_0$ is the mass
of the particles, and $\mathbf{A}(\mathbf{x})$ is a slowly-varying vector potential. 
The resultant free-energy density has the standard GL form~\cite{degennes66, stintzing97, iskin11a}
\begin{align}
\mathcal{F}_\mathrm{GL} = \mathcal{F}_\mathrm{n} &+ \alpha(T) |\Psi|^2 
+ \frac{1}{2 m_0} \left|\left( -\mathrm{i} \hbar \boldsymbol{\nabla} 
- \frac{2q_0}{c_0} {\bf A} \right) \Psi\right|^2	
\nonumber \\
&+ \frac{\beta_0}{2} |\Psi|^4
+ \frac{|\mathbf{H}|^2}{8\pi},
\label{eqn:freeE}
\end{align}
where the $\mathbf{x}$-dependences are suppressed. While 
$
\alpha(T) = \hbar^2 a(T)/(m_0 c)
$ 
is negative for $T < T_c$ and it changes sign at $T = T_c$, a positive 
$
\beta_0 = \hbar^4 \mathcal{V}_\mathrm{uc} b_0/(m_0 c)^2
$ 
(or equivalently $b_0$) guarantees the energetic stability of the theory. 
Here $\mathcal{V}_\mathrm{uc}$ is the volume of the unit cell. 
In principle $m_0$ is arbitrary in the GL theory, and physical quantities do not depend on 
it~\cite{degennes66, leggett}. 

In Sec.~\ref{sec:lee} we showed that quantum geometry of the underlying Bloch states has 
a partial control over those thermodynamic properties that has explicit dependence on 
the kinetic coefficient $c$. For instance the London penetration depth
$
\lambda = \sqrt{ m_0 c_0^2/(16\pi q_0^2 |\Psi|^2) }
$
is one of them~\cite{degennes66, stintzing97, iskin11a}, where we substitute
$
|\Psi|^2 = -\alpha(T)/\beta_0 
$ 
assuming weak magnetic fields. This leads to
$
\lambda (T) = \lambda_\mathrm{GL} / \sqrt{\epsilon (T)},
$
where 
\begin{align}
\label{eqn:lambda}
\lambda_\mathrm{GL} = \sqrt{\frac{\hbar^2 \mathcal{V}_\mathrm{uc} b_0}{16\pi m_0 r_0 a_0 c}}
\end{align}
is a temperature-independent prefactor. Here $r_0 = q_0^2/(m_0 c_0^2)$ is the classical 
radius of a particle with mass $m_0$ and charge $q_0$ in CGS units. 
Similarly the GL coherence length
$
\xi(T) = \hbar/\sqrt{2 m_0 |\alpha (T)|}
$
is another example~\cite{degennes66, stintzing97, iskin11a}, leading to 
$
\xi (T) = \xi_\mathrm{GL} / \sqrt{\epsilon (T)}
$
where 
\begin{align}
\label{eqn:xi}
\xi_\mathrm{GL} = \sqrt{\frac{c}{2 a_0}}
\end{align}
is a temperature-independent prefactor. Their ratio 
$
\kappa = \lambda (T)/\xi (T)
$ 
is known as the GL parameter, and it determines whether the superconductor
is of type-I or type-II depending on whether $\kappa < 1/\sqrt{2}$ or 
$\kappa > 1/\sqrt{2}$, respectively. This parameter also depends on quantum geometry 
through
\begin{align}
\label{eqn:kappa}
\kappa = \sqrt{\frac{\hbar^2 \mathcal{V}_\mathrm{uc} b_0}{8\pi m_0 r_0 c^2}},
\end{align}
and it is independent of temperature. In addition the upper critical field for type-II
superconductors 
$
H_{c_2} (T) = \Phi_0/[2\pi \xi^2(T)]
$
is another example~\cite{degennes66, stintzing97, iskin11a}, where 
$
\Phi_0 = \pi \hbar c_0/q_0
$ 
is the superconductivity flux quantum. This leads to
$
H_{c_2} (T) = H_{c_2}^0 \epsilon (T), 
$
where 
\begin{align}
H_{c_2}^0 = \sqrt{\frac{a_0^2}{m_0 r_0 c^2}}
\end{align}
is a temperature-independent prefactor.

Furthermore the superfluid number density $\rho_\mathrm{sf}(T)$ can be obtained from the 
GL free energy of the current-carrying superfluid as 
follows~\footnote{See also Ref.~\cite{leggett} where the superfluid mass density is 
derived for a generic GL functional in page 200, or the discussion in 
Ref.~\cite{stintzing97}}. 
Under the assumption of a spatially-uniform condensate, one first imposes the 
current by applying a phase twist
$
\Psi(\mathbf{x}) = |\Psi|e^{\mathrm{i} \mathbf{q}_\mathrm{sf} \cdot \mathbf{x}}
$
to the order parameter and reexpress
$
|\boldsymbol{\nabla} \Psi(\mathbf{x})|^2 = q_\mathrm{sf}^2 |\Psi|^2 
$
in Eq.~(\ref{eqn:freeE}), where $\mathbf{v}_\mathrm{sf} = \hbar \mathbf{q}_\mathrm{sf}/(2m_0)$ 
is the superfluid velocity associated with the imposed twist. Then $\rho_\mathrm{sf}(T)$
is determined from the extra free-energy density
$
\Delta \mathcal{F} = m_0 \rho_\mathrm{sf}(T) v_\mathrm{sf}^2/2
$
of the imposed superfluid flow, where $m_0 \rho_\mathrm{sf}(T)$ is the so-called superfluid 
mass density. As a result we identify 
$
\rho_\mathrm{sf}(T) = 4 |\Psi|^2,
$  
leading to
\begin{align}
\label{eqn:rhos}
\rho_\mathrm{sf}(T) = \frac{4m_0 a_0 c}{\hbar^2 \mathcal{V}_\mathrm{uc} b_0} \epsilon (T).
\end{align} 
Note that Eq.~(\ref{eqn:rhos}) of the GL theory does not coincide with the well-established 
result that is based on the more appropriate linear-response theory~\cite{liang17},
and it may lead to qualitatively accurate but quantitatively inaccurate results in the 
strong-coupling limit, e.g., see~\footnote{For instance Eq.~(\ref{eqn:rhos}) is used in 
Ref.~\cite{stintzing97} to calculate $T_\mathrm{BKT}$ for a two-dimensional continuum 
Fermi gas with contact interactions. In contrast to the well-established and physically-expected 
result $T_\mathrm{BKT} = 0.125\varepsilon_F$~\cite{loktev01}, they found that the BKT 
transition temperature saturates around $T_\mathrm{BKT} \sim 0.11\varepsilon_F$ in the 
strong-coupling limit, where $\varepsilon_F$ is the Fermi energy. This is clearly a 
qualitatively accurate but a quantitatively inaccurate result.} and Sec.~\ref{sec:aif}.
Since the phase stiffness $D_\mathrm{sf}(T)$ (also known as the helicity modulus or 
the superfluid weight) is given by 
$
D_\mathrm{sf}(T) = \hbar^2 \rho_\mathrm{sf} (T) /m_0
$
~\cite{iskin18c}, the universal Beresinskii-Kosterlitz-Thouless (BKT) relation 
$
T_\mathrm{BKT} = \pi D_\mathrm{sf} (T_\mathrm{BKT})/8,
$
which determines the superfluid transition temperature in two dimensions, 
can be written as
\begin{align}
T_\mathrm{BKT} = \frac{\pi a_0 c}{2\mathcal{A}_\mathrm{uc} b_0} \epsilon(T_\mathrm{BKT}),
\label{eqn:BKT}
\end{align}
where $\mathcal{A}_\mathrm{uc}$ is the area of the unit cell.
Thus effective mass of the Cooper pairs, London penetration depth, GL coherence length, 
GL parameter, upper critical magnetic field, superfluid density and BKT transition temperature 
are some of those fundamental thermodynamic properties that are partially controlled 
by the quantum-geometric effects.

\subsection{Application to an isolated flat band}
\label{sec:aif}

As an application here we consider a weakly-coupled dispersionless flat band that is 
energetically isolated from the rest of the Bloch bands in the spectrum~\cite{peotta15}. 
Suppose 
$
\varepsilon_{f \mathbf{k}} = \varepsilon_f
$ 
is the energy of this flat band, and it is separated by an energy $\varepsilon_0$ from 
the nearest band. One can show that 
$
\mu = \varepsilon_f - U(1-F_f)/(2N_S)
$ 
in the $U/\varepsilon_0 \to 0$ limit~\cite{iskin17}, where $0 \le F_f \le 2$ is the 
band filling~\footnote{
When $F_f \to 0$, we recognize that the binding energy of the lowest bound state 
is $U/N_s$ in the $U/\varepsilon_0 \to 0$ limit~\cite{iskin21, iskin22}.
}, 
and we expand 
$
\mathcal{X}_{f\mathbf{k}} / \xi_{f \mathbf{k}} 
\approx \beta/2 - \beta^2 \xi_{f \mathbf{k}}/24 + \cdots
$
and 
$
\mathcal{Y}_{f\mathbf{k}} \approx 1 - \beta^2 \xi_{f \mathbf{k}}^2/4 + \cdots.
$
This leads to $a_0 = 1/(4T_c)$, $b_0 = 1/(48T_c^3)$,
$
d = [1/(\varepsilon_f - \mu) + \mathrm{i} \pi \delta(\varepsilon_f - \mu)]/(8 T_c),
$
and
$
c = \frac{1}{4T_c N_c} \sum_\mathbf{k} g_0^{f \mathbf{k}},
$
where $g_0^{f \mathbf{k}}$ is the isotropic quantum metric of the flat band. 
See~\cite{sademelo93} for a comparison with the dispersive-band case.
Since $d$ diverges as $\mu \to \varepsilon_f$ in the $U/\varepsilon_0 \to 0$ limit, 
the Cooper pairs have large effective mass, they are weakly repulsive with each other, 
and their size is small compared to the inter-particle spacing. This is because, 
given that $U/W \gg 1$ with $W \to 0$ the band width of the 
flat band, even an arbitrarily small but finite $U$ corresponds effectively to a 
strong-coupling limit. In addition while the Thouless condition gives $T_c = U/(4N_S)$, 
Eq.~(\ref{eqn:BKT}) leads to
$
T_\mathrm{BKT} \approx \frac{3\pi U}{8N_S \mathcal{A}} \sum_\mathbf{k} g_0^{f \mathbf{k}},
$
which is proportional to the inverse-effective mass
$
\frac{U}{\hbar^2 N_S N_c} \sum_\mathbf{k} g_0^{f \mathbf{k}}
$
of the two-body bound states~\cite{torma18, iskin21}. 
Here $\mathcal{A} = N_c \mathcal{A}_\mathrm{uc}$ is the area of the system,
and we assumed $T_\mathrm{BKT}  \ll T_c$.
Note that the BKT transition temperature that follows from our simple GL theory almost
coincides with the literature, where the prefactor is known to be $1/4$ instead 
of $3/8$ at half filling~\cite{peotta15}. The root cause of this discrepancy is due to
the inaccurate form of Eq.~(\ref{eqn:rhos}) in the strong-coupling limit. 
Thus, as a result of our GL analysis, we conclude that $\xi_\mathrm{GL}$ and $H_{c_2}^0$ 
are independent of $U$ for a flat-band superconductor when $U/\varepsilon_0 \to 0$, 
because the ratio 
$
c_{ij}/a_0 = \frac{1}{N_c} \sum_\mathbf{k} g_{ij}^{f \mathbf{k}}
$
depends only on the average quantum metric over the Brillouin zone.

\section{Conclusion}
\label{sec:conc}

In summary, to gain insight into the thermodynamic properties of multiband superconductors, 
here we considered a generic multiband Hubbard model, and derived its effective quartic 
action in the fluctuations of the pairing order parameter near $T_c$. 
The effective action is remarkably simple when the system manifests time-reversal 
symmetry and it exhibits uniform pairing fluctuations in a unit cell at both low 
momentum and low frequency. In this particular case we derived the TDGL equation, 
and showed that its kinetic coefficients have a geometric interband contribution that 
is controlled by both the quantum-metric tensor of the underlying Bloch states and 
their band-resolved quantum-metric tensors. Through a standard GL analysis, 
we extracted examples of thermodynamic properties such as the effective mass of the
Cooper pairs, London penetration depth, GL coherence length, GL parameter, upper critical 
magnetic field, superfluid density and BKT transition temperature which have explicit 
dependence on quantum geometry. Looking ahead our analysis can be extended to the 
broken-symmetry state at temperatures 
$T \ll T_c$~\cite{engelbrecht97}, i.e., by again assuming 
time-reversal symmetry and uniform pairing fluctuations at both low momentum and 
low frequency. This would verify that the same quantum-geometric terms control the 
velocity of the Goldstone modes in a completely analogous way~\cite{iskin20a, iskin20b}.

\textit{Note added}. While finalizing this manuscript, Ref.~\cite{chen23} appeared 
in the preprint server, where GL theory is derived for an isolated flat band.

\begin{acknowledgments}
The author acknowledges funding from T{\"U}B{\.I}TAK.
\end{acknowledgments}

\bibliography{refs}

\end{document}